# Energy Efficient Skyrmion based Oscillator on Thermocoupled Nanotrack

Ravish Kumar Raj[1*], Namita Bindal[1], and Brajesh Kumar Kaushik[1]

[1] *Department of Electronics and Communication Engineering,*
*Indian Institute of Technology Roorkee, Roorkee, India 247667*

The magnetic skyrmion-based spin transfer nano-oscillators (STNO) are the potential candidates for next-generation microwave signal generator and has gained popularity due to their performance, integrability and compatibility with existing CMOS technology. However, these devices suffer from the Joule heating problem that neglects their non-volatility advantage in spintronic devices. Therefore, it is necessary to investigate the alternative driving mechanisms for the development of energy-efficient skyrmion based nano-oscillators. In this paper, a skyrmion-based nano-oscillator has been designed that utilizes thermal power to drive skyrmion on a thermocoupled nanotrack. The thermocoupled nanotrack is designed in such a way that both the upper and lower nanotracks have different values of damping constants and a temperature difference is maintained between the extreme ends, in order to create a temperature gradient in the two nanotracks. By employing this technique, skyrmion is able to exhibit the periodic motion on the nanotrack with the maximum achievable frequency of 2.5GHz without any external stimuli. Moreover, the proposed device offers low thermal energy consumption of 0.84fJ/oscillation. Hence, this work provides the pathway for the development of energy-efficient future spintronic devices.

## I. INTRODUCTION

Magnetic skyrmion is a topologically stable chiral spin texture that has a localized excitation of magnetic moments and exists in magnetic materials as a result of Dzyaloshinski-Moriya interaction (DMI) that is associated with broken inversion symmetry and strong spin-orbit coupling [1-4]. Recently, ultra-small (100ns to 1ns) sized skyrmions with a few nanometers dimension have been reported in ultrathin ferromagnetic (FM) and antiferromagnetic (AFM) materials [5]. Later, the skyrmions were observed at the room temperature (RT) in magnetic multilayers like Ta/CoFeB/TaOx and Pt/CoFeB/MgO. Owing to low current density requirements and nanoscale size, the magnetic skyrmions are attracting much interest both in fundamental science and applications in novel devices, especially for non-volatile magnetic memories [6], logics [7], transistors [8], diodes [9], oscillators [10], and neuromorphic computing [11].

In the recent past, very small-sized spin-based oscillators such as spin transfer nano-oscillator (STNO) and spin Hall nano-oscillator (SHNO) have been designed however, these devices limit the scalability and tenability [12-13]. Hence, in view of these shortcomings, the skyrmion-based oscillators that employ vortex-like polarization at the fixed layer of magnetic tunnel junction (MTJ) were proposed [14-15]. Moreover, these are driven by the spin-polarized current that includes spin transfer torque (STT) and spin-orbit torque (SOT) mechanisms that requires a large current and subsequently translate into a large energy dissipation (~100fJ). This is at least 3 orders of magnitude larger than the energy dissipation in CMOS-based oscillators (~100aJ) [16]. Therefore, these devices offer the problem of Joule heating and this neglects the advantage of non-volatility offered by spintronic devices. Hence, there is an urgent need for an alternative driving mechanism to designing skyrmion based nano-oscillators. The alternatives include magnetic anisotropic gradient, DMI gradient, strain gradient as well as temperature gradient (TG). Out of these mechanisms, for the practical application, the TG is the best choice for driving the skyrmions due to its low energy consumption, hence TG driven skyrmion is the basis of a new research direction named as skyrmion-caloritronics [17]. In addition, it contributes to the design of waste heat recovery that can be further utilized for nucleation and manipulation of the skyrmion. However, the investigation behind the dynamics of skyrmion under TG is still in its early stage.

In recent study, it has been found that the dynamics of skyrmion in the presence of TG can be attributed to the competition among different torques and fields including, the magnonic torque, thermal torque, entropy difference, and thermally induced dipolar field [18-19]. However, the experimental results have shown contradictory outcomes, with skyrmions either moving towards colder regions or hotter regions [20]. The magnonic torque that is a result of the propagation of magnons from hotter to colder regions, exerts a spin torque on the skyrmion, pushing it towards the hotter region to minimize the system's free energy [21].

Additionally, the entropy equivalent field, adiabatic torque for positive chirality, and thermally induced dipolar field also drive the skyrmion towards the hotter region [22]. Conversely, other effects, such as force due to thermal torque and entropy difference, push the skyrmion toward the colder region [23]. Ultimately, the net motion of the skyrmion depends on the relative strength of these competing factors that is determined by material parameters as well as the magnetic structure [20].

In this work, a skyrmion-based nano-oscillator is designed by connecting the two nanotracks (upper (*A*) and lower (*B*) semi-circle nanotrack) with different damping constants ($\alpha_2$ and $\alpha_1$) end-to-end in a circular manner, similar to a thermocouple. The temperature of the two extreme ends is maintained at $T_1$ and $T_2$ to employ the radial TG on nanotrack that led to the complete periodic motion i.e. oscillation of the skyrmion motion. The micromagnetic simulations have been performed to determine the required magnitude of radial TG, the width of the nanotracks, and the damping constant difference between the two nanotracks in order to achieve proper oscillatory motion. Additionally, the maximum frequency of the oscillator has also been determined. The proposed device offers a promising path for the development of energy-efficient non-volatile devices with ultra-high storage density.

## II. MICROMAGNETIC MODELLING

In this work, Mumax3, a GPU-accelerated micromagnetic simulation tool is used to describe finite cell difference of the magnetization field by a nonlinear Landau Lifshitz-Gilbert (LLG) equation that is defined as follows [24-25]:

$$\frac{dm}{dt} = -\frac{\gamma_0}{1+\alpha}\left[m \times (H_{eff} + h_{th}) + \alpha\left(m \times (m \times (H_{eff} + h_{th}))\right)\right] + \tau_{th} \quad (1)$$

where $\alpha > 0$ and $\gamma_0 > 0$ are the Gilbert damping coefficient and gyromagnetic ratio, respectively. ***m***=***M***/$M_s$ is the normalized magnetization vector with respect to saturation magnetization $M_s$. The net effective field is defined as $\boldsymbol{H_{eff}} = -\frac{1}{\mu_0 M_s}\frac{\partial \varepsilon}{\partial \boldsymbol{m}}$ and micromagnetic energy density $\varepsilon$ is expressed as follows [26]:

$$\varepsilon = \varepsilon_{exch} + \varepsilon_{anis} + \varepsilon_{DMI} + \varepsilon_{demag} \quad (2)$$

$$\varepsilon = A_{ex}[(\nabla \mathbf{m})^2] - K_u(\boldsymbol{u}.\boldsymbol{m})^2 + D_0[m_z(\nabla.\boldsymbol{m}) - (\boldsymbol{m}.\nabla)m_z] - \frac{\mu_0 M_s \boldsymbol{m}.\boldsymbol{H_{demag}}}{2} \quad (3)$$

The net effective field $H_{eff}$ is a contribution of various energies, such as symmetric exchange energy ($A_{ex}$ as exchange constant), uniaxial crystalline anisotropy energy ($K_u$ as an anisotropic constant), non-centrosymmetric DMI energy ($D_0$ as a DMI constant), and self-demagnetization energy with demagnetizing field $H_{demag}$. The suggested device is simulated by computing LLGS equation based on the experimental material parameters at zero temperature that are listed in Table 1. Moreover, as this work is based on TG, the thermally induced Gaussian stochastic field ($h_{th}$) is also a contributing factor in net effective field calculation. The autocorrelation related to this is given as follows [29-30]:

$$\langle h_{th,ir}(t) h_{th,js}(t + \Delta t) \rangle = h_{th}\delta_{ij}\delta_{rs}\delta(\Delta t) \quad (4)$$

where $\boldsymbol{h_{th}} = \xi\sqrt{\frac{2\alpha k_B T}{\mu_0 M_s \gamma_0 V \Delta t}}$ and $\xi$ is a six-dimensional random variable with white Gaussian noise having mean and variance as 0 and 1, respectively. *r* and *s* denote the effective Cartesian components of the induced thermal field. $k_B$ and $\Delta t$ are the Boltzmann constant and time interval, respectively. However, *T* and *V* depict the temperature and volume of the computational cell, respectively. The overall dynamics of the magnetization with the thermal perturbation are described by considering the effects of potential thermal spin-transfer torques ($\tau_{th}$), that include the adiabatic and non-adiabatic STTs in the stochastic LLG equation (1) and is defined as follows [18]:

$$\boldsymbol{\tau_{th}} = u\boldsymbol{m} \times \left(\frac{d\boldsymbol{m}}{dx} \times \boldsymbol{m}\right) - u\beta\left(\boldsymbol{m} \times \frac{d\boldsymbol{m}}{dx}\right) \quad (5)$$

where, $u = \left|\frac{\gamma_0 \hbar}{e \mu_0}\right| \frac{j_s P}{M_s(1+\beta^2)}$ is the STT coefficient induced by the thermal current. $\beta$ is the degree of non-adiabaticity. However, $\mu_0$, $\hbar$, $e$, $j_s$, and $P$ are the reduced Planck constant, absolute permeability, electron charge, spin current density, and spin polarization factor, respectively. The effects of temperature on material parameters are explained in detail in Supplemental Material 1.

Table 1. Material parameters of the proposed device for micromagnetic simulations at 0K [27-28]

| Parameters | Values |
|---|---|
| Saturation Magnetization $M_S$ | $580 \, kA/m$ |
| Exchange Stiffness $A_{ex}$ | $15 \times 10^{-12} \, J/m$ |
| Spin polarization rate $P$ | 0.6 |
| Non-adiabatic STT $\beta$ | 0.3 |
| Interfacial DMI $D_{ind}$ | $3 \times 10^{-3} \, J/m^2$ |
| Anisotropic Constant $K_u$ | $8.5 \times 10^5 \, J/m^3$ |
| Elementary Charge $e$ | $1.6 \times 10^{-19} \, C$ |

## III. DYNAMICS OF SKYRMION IN TEMPERATURE GRADIENT

The behavior of skyrmion in TG is studied using micromagnetic simulations and it shows controversial dynamics with respect to the propagation direction (i.e., whether the skyrmion moves towards the colder or hotter side). This behavior is influenced by various factors, including magnetic structure and material parameters. To analyze this, different types of forces that act on the skyrmion, namely magnonic torque ($F_{\mu STT}$), thermal torque ($F_{\tau STT}$), and dipolar field ($F_{DF}$) are considered that are explained using various macroscopic thermodynamic theories, including the spin-dependent Peltier effect [31], spin Nernst effect [32], thermal Hall effect [33], and spin Seebeck effect [34].

For this controversial dynamics of skyrmion, analysis is done using micromagnetic simulations to investigate the skyrmion behavior under TG on a straight nanotrack by considering the forces experienced by the skyrmion due to magnonic torque ($F_{\mu STT}$), thermal torque ($F_{\tau STT}$), and dipolar field ($F_{DF}$) that is defined as follows:

$$\boldsymbol{F_{TG\_net} = F_{\mu STT} + F_{\tau STT} + F_{DF}} \quad (6)$$
$$\boldsymbol{F_{ext} = F_{SkHE} + F_{sky-edge}} \quad (7)$$

From equation (6), it is concluded that the net motion of the skyrmion in the longitudinal direction will depend upon the relative effects of forces due to magnonic STT, thermal STT, and the dipolar field induced by TG with thermally induced drag force acting opposite to the motion of skyrmion. However, equation (7) depicts that the motion of the skyrmion in the radial direction is induced by the thermal Hall effect, that is further counter balanced by the repulsive force due to the enhancement of the potential energy of the skyrmion near the edge. The direction of skyrmion motion under the effect of different torques is shown in Fig. 1. and is discussed in detail in the following sub-sections.

### A. Magnonic torque ($F_{\mu STT}$)

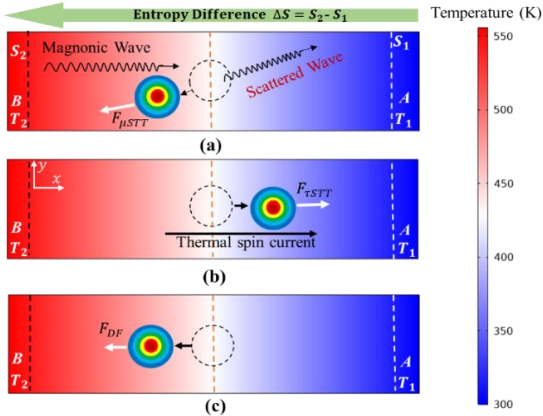

FIG. 1. Forces acting on the skyrmion due to variation of temperature from $T_2$ to $T_1$ ($T_2 > T_1$) along a nanotrack (a) Magnonic torque, that acts due to the skew scattering of magnonic wave (generated by entropy difference along the length) by skyrmion through transfer of angular momentum, magnonic torque ($F_{\mu STT}$) acts on the skyrmion and moves it in the direction opposite to the propagation of spin wave (b) Thermal torque, the TG on a nanotrack generates pure spin current (spin Seebeck effect) that exerts the thermal torque ($F_{\tau STT}$) and it pushes the skyrmion towards the colder side (c) Dipolar field, that is thermally induced along the length, drives the skyrmion from colder to hotter side.

Under TG, random fluctuations (conical oscillation) of spins about its mean position are larger in the region where temperature is highest on the nanotrack and varies along the length in proper phase difference and this results in the generation of magnonic wave in the hotter side ($T_2$) that starts propagating towards the colder one ($T_1$). This is due to the fact that the significant amount of temperature difference at the nanotrack ends leads to the entropy difference ($S_2 - S_1$) along the length [19]. As soon as the magnonic wave reaches the skyrmion, it interacts with the spin of the skyrmion and gets scattered in an oblique direction and it leads to the transfer of angular momentum between them thus, pushing the skyrmion towards the hotter side by exerting $F_{\mu STT}$ as shown in Fig. 1(a). However, the magnonic wave is decaying during the propagation along the length due to the damping effect of spins that directly depends on the damping constant ($\alpha$) [35]. Hence, $F_{\mu STT}$ decays rapidly along the length of the nanotrack for the larger value of $\alpha$ as compared to the nanotrack having smaller damping constant.

### A. Thermal torque ($F_{\tau STT}$)

The different energy states are created owing to the TG on the nanotrack and this leads to the diffusion of up and down electrons in opposite directions and due to these conduction electrons have different Seebeck coefficients [36]. Hence, the spin current is generated that flow toward the colder side and exerts $F_{\tau STT}$ on the skyrmion through the STT effect that pushes the skyrmion towards the colder side of the nanotrack as shown in Fig. 1(b). However, the magnitude of thermal spin current depends upon the electrical conductivity and spin Seebeck coefficient that is inversely proportional to damping constant ($\alpha$), meaning the spin Seebeck effect is stronger for low damping, that results in larger spin current to flow from hotter to colder [37]. This means that the lower damping constant has stronger $F_{\tau STT}$ and permits to move the skyrmion towards the colder side.

### B. Thermally induced dipolar force ($F_{DF}$)

Under TG on a nanotrack due to random fluctuations, dipolar field is induced that affects the dynamics of skyrmion. The change in the dipolar field along the nanotrack produces a stray field gradient that pushes the skyrmion towards the hotter side (shown in Fig. 1(c)) by exerting $F_{DF}$ as there is stronger dipolar field at larger temperature [19]. Moreover, larger damping constant have more tendency to align the spin to the effective field that reduce the induced dipolar field due stray field have less interaction with spins.

## IV. THIELE'S EQUATION FOR SKYRMION ON A CIRCULAR NANOTRACK

It is assumed that the skyrmion is an elastic rigid body and the dynamics of skyrmion is defined by the Thiele equation on solving the LLG equation by considering only the lowest energy excitations of the skyrmions [18]. The skyrmion motion is modelled using the Thiele equation to be extended by an additional term that is a scalar quantity, magnon induced friction ($\xi T$), in order to include the effect of thermal fluctuations [38]. In this work, the skyrmions motion is along the circular track (in the form of thermocouple), and hence the Thiele equation has been derived in terms of angular velocity $\omega$ that is expressed as follows:

$$M_{eff}(\boldsymbol{a_t} \times \boldsymbol{R_{eff}}) + \boldsymbol{G_0} \times (\boldsymbol{\omega} \times \boldsymbol{R_{eff}}) - (D_0\alpha + \xi T)(\boldsymbol{\omega} \times \boldsymbol{R_{eff}}) = \boldsymbol{F_{TG}} \quad (8)$$

where $M_{eff}$ and $R_{eff} = \frac{R_1+R_2}{2}$ are the effective mass and radius of motion on nanotrack of the skyrmion, respectively. $a_t$ denotes the tangential acceleration. $G_0 = (0,0,-4\pi Q)$ is the gyromagnetic vector that depends on the topological charge that is defined as $Q = (1/4\pi)\int\{\boldsymbol{m}\cdot(\partial_x\boldsymbol{m}\times\partial_y\boldsymbol{m})\}dxdy$, where $m$ is the unit magnetization vector [19]. This charge is quantized to an integer number (either -1 or +1) and it depends on the core polarity i.e. $p = [m_z(r=\infty) - m_z(r=0)]/2$ of the skyrmion [39]. $D_0$ is the dissipative tensor that is given as $D_0 = \begin{pmatrix} D' & 0 \\ 0 & D' \end{pmatrix}$ with $D' = \int dxdy\delta_x\boldsymbol{m}\cdot\delta_x\boldsymbol{m} = \int dxdy\delta_y\boldsymbol{m}\cdot\delta_y\boldsymbol{m}$ [27]. On solving the vector algebra of equation (8), the frequency of the skyrmion after attaining the uniform circular motion is defined as follows:

$$f = \frac{1}{2\pi R_{eff}} \frac{F_{TG}}{\sqrt{(G_0)^2 + \{(D_0\alpha + \xi T)\}^2}} \quad (9)$$

Here, $F_{TG}$ is the force due to TG in the effective field along the $xy$ plane. The detailed derivation is discussed in Supplemental Material 2.

## V. SCHEMATIC AND OPERATION OF THE PROPOSED DEVICE

The proposed device is a thermocouple that mimics as skyrmion based oscillator and it formed by connecting two FM nanotracks $A$ and $B$ end to end in a circular manner with dissimilar damping constants $\alpha_2$ and $\alpha_1$, respectively, as shown in Fig. 2. The TG is created between the two ends in such a way that the left most junction $J_2$ i.e. $f(x_0, y_0) = (-200, 0)nm$ is at the highest temperature ($T_2 = 540K$) whereas right most junction $J_1$ i.e. $f(x_0, y_0) = (200, 0)nm$ is at room temperature, $T_1 = 300K$. The inner radius and outer radius of nanotrack are considered as $R_2 = 250nm$ and $R_1 = 150nm$, respectively in order to achieve the periodic motion of skyrmion. On the circular nanotrack, there is a linear variation of temperature (300 to 540K) with angular displacement ranging from $\theta = 0°$ to $\theta = \pm 180°$ and it is expressed as $T(R_{eff}, \theta) = T_1(R_{eff}, 0°) + (\nabla T \cdot R_{eff})\theta$, where $\nabla T = \frac{1}{w_n}\frac{\partial T}{\partial \theta}$ is the radial TG and $\partial T = T_2 - T_1$. Here, $\partial \theta$ is the angular displacement. However, $w_n = R_2 - R_1$ is the width of nanotrack.

The proposed device operates as an oscillator with the skyrmion performing periodic motion on the nanotrack. Basically, the skyrmion undergoes repetitive motion on the nanotrack, means reaching the same position after a definite time period $T$. For this, initially, the skyrmion is nucleated at the point $f(x_0, y_0) = (-200, 20) nm$ with the help of MTJ $M_2$ near to junction $J_2$ on nanotrack $A$. Thereafter, the system is relaxed for 200ps, and then the skyrmion started moving in counter clockwise (CCW) direction toward the nanotrack $B$ due to the fact that the force caused by thermal STT dominates over the other two forces including magnonic torque and dipolar field. The thermal spin current density ($T_2$ to $T_1$) is more in nanotrack $B$ as compared to nanotrack $A$ since the damping constant of nanotrack $B$ is lower than that of nanotrack $A$. Afterward, the skyrmion moves toward the $J_1$ through the nanotrack $B$. As soon as it reaches the junction point $J_1$, the skyrmion enters the nanotrack $A$ due to inertial effect of skyrmion. As the damping constant of the nanotrack $A$ is higher as compared to nanotrack $B$, the force caused by magnonic STT and induced dipolar field dominates over the force due to thermal STT, hence the skyrmion experiences the net effective force toward the hotter side. The skyrmion now moves toward the $J_2$ in the CCW direction through nanotrack $A$ and finally crosses the junction $J_2$ and reaches nanotrack $B$ due to inertial effect. In this way, the complete periodic motion of the skyrmion in the CCW direction is achieved. This motion translates into a spike signal whenever it is detected by the MTJ detector and is referred to as the oscillator signal. This signal can be further manipulated using an RC circuit. For detection of skyrmion, the two MTJs that are placed at $M_2 = f(x_0, y_0) = (-200, 20) nm$ and $M_1 = f(x_0, y_0) = (200, -50) nm$ with respect to origin (O) for out of phase (positive and negative spike) signal detection with a proper phase difference.

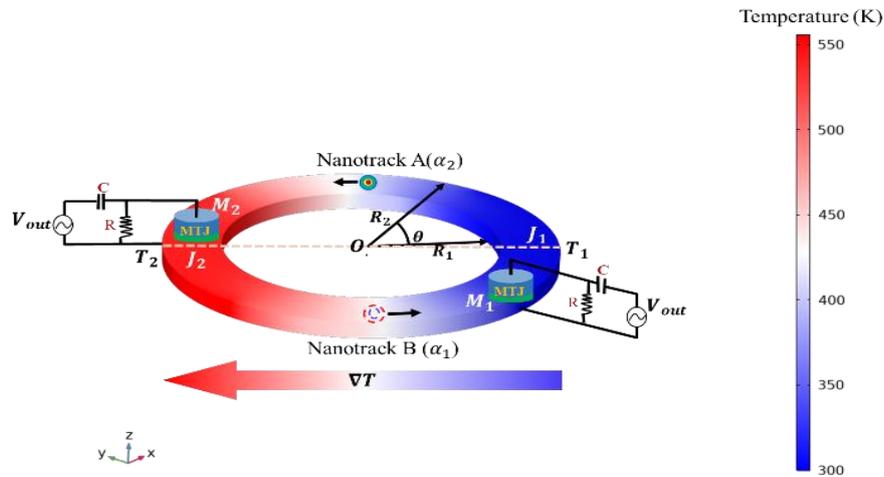

FIG. 2 A thermocoupled nanotrack with radial TG (where $T_2 > T_1$) along the circular nanotrack. Two nanotracks $A$ and $B$ have different damping constants $\alpha_2$ and $\alpha_1$, respectively, where $\alpha_2 > \alpha_1$.

Fig. 3(a) shows the various forces that are acting on the skyrmion during its motion on the nanotrack including $F_{TG}$, $F_{drag}$, $F_{sky-edge}$, and $F_{SkHE}$ that are due to radial TG, magnon induced friction, skyrmion-edge interaction and skyrmion Hall effect (SkHE), respectively. SkHE is responsible for the deviation of the skyrmion from the straight path that can further lead to the annihilation of the skyrmion at the boundaries. The dynamics of skyrmion in the presence of TG are affected by the forces experienced by the skyrmion in the tangential direction due to magnonic torque ($F_{\mu STT}$), thermal torque ($F_{\tau STT}$), and dipolar field ($F_{DF}$). However, $F_{drag}$ acts in the opposite direction to that of $F_{TG}$ in order to include the effect of thermal fluctuations. Due to SkHE, the skyrmion will approach near the edge of nanotrack resulting in enhanced potential energy of the skyrmion that induces repulsive force $F_{sky-edge}$. However, $F_{sky-edge}$ and $F_{SkHE}$ are acting in the opposite radial direction thereby, protecting the skyrmion from any annihilation and deformation. Hence, the skyrmion is confined in the potential wall due to the presence of high potential at both the inner and outer edges of the nanotrack. As a result, the skyrmion experiences a repulsive force that pushes the skyrmion toward the centre of nanotrack whenever it reaches either side of the edge. All these forces are responsible for the proper circular motion of the skyrmion and its time evolution of position in $xy$ plane is shown in Fig. 3(b).

Under the presence of TG, the forces acting on the skyrmion are strongly dependent on the damping constant. Hence, in order to design skyrmion based oscillator on a thermocoupled nanotrack, there is a need to investigate the difference in values of the damping constants that are considered for the two nanotracks $A$ and $B$ (connected end-to-end). Moreover, radial direction of net force acting on the skyrmion due to SkHE and from the nanotrack edges play a crucial role in deciding whether the skyrmion will be able to perform periodic motion without any deformation/annihilation that depends on width of the nanotrack ($w_n$) and the difference between the damping constant ($\alpha_2 - \alpha_1$). Therefore, the working windows illustrating the period motion of skyrmion the for width of nanotrack and damping constant difference under a wide range of radial TG is discussed in the next section.

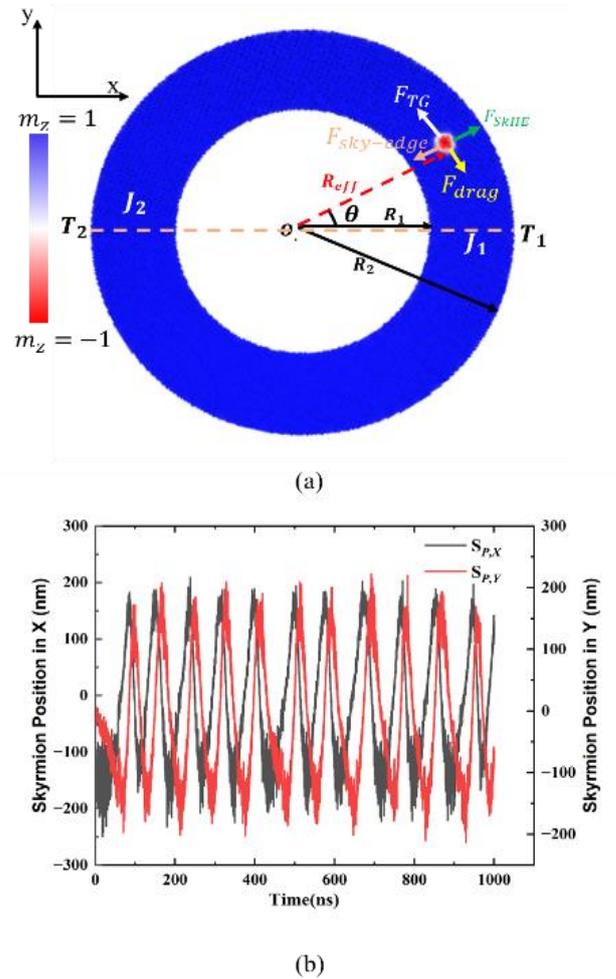

FIG. 3. (a) Forces exerted on the skyrmion during its motion on circular nanotrack. $F_{TG}$ and $F_{drag}$ are the forward and backward forces that act in tangential direction, whereas the gyrotropic force $F_{SkHE}$ and $F_{sky-edge}$ acts in radial direction to nanotrack. (b) Time evolution of skyrmion position in $xy$ plane for $\alpha_2 - \alpha_1 = 0.004, w_n = 200 nm$ and RTG is 10 K/$\mu$m-degree.

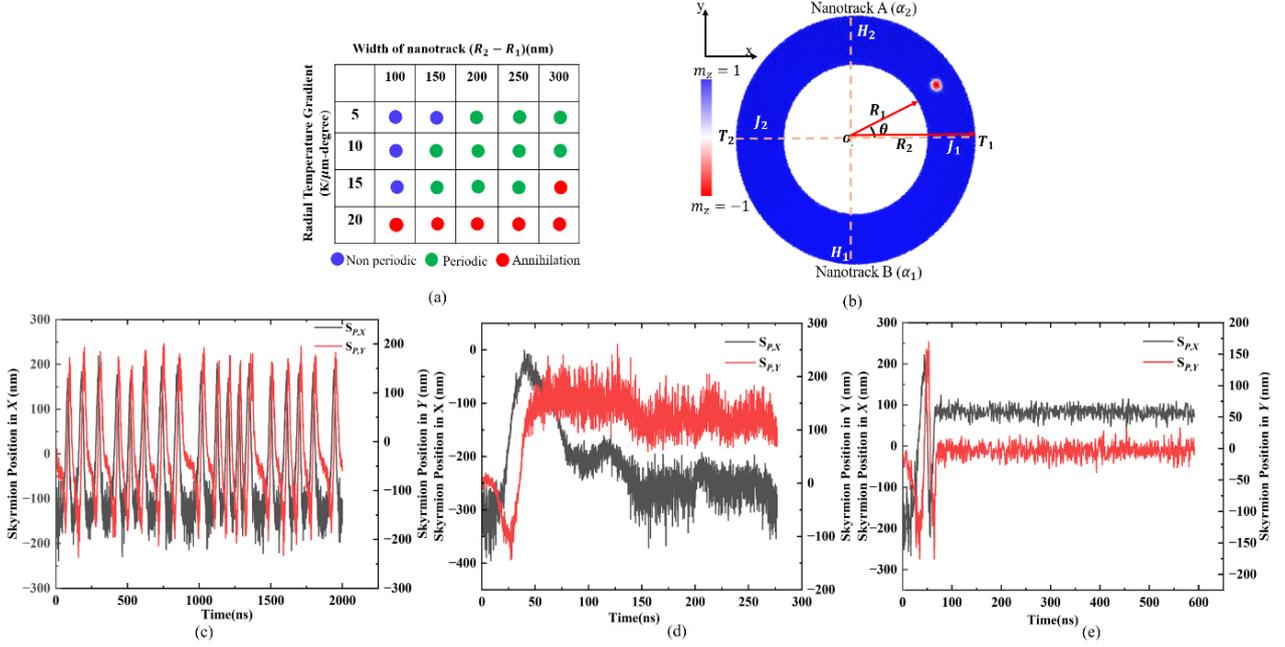

FIG. 4. (a) Working window for different skyrmion behavior on a circular nanotrack for width $100 \leq w_n \leq 300$ and $5 \leq \nabla T \leq 20$ K/μm-degree with $\alpha_2 - \alpha_1 = 0.004$ (b) Illustration of different points on nanotrack where different forces are acting on the skyrmion (c) Instantaneous skyrmion position when it exhibits periodic motion (d) Non-periodic motion and (e) Annihilation on a circular nanotrack.

## VI. WORKING WINDOW OF THE PROPOSED DEVICE FOR THE NANOTRACK WIDTH

The working window for a circular motion of skyrmion on a nanotrack is shown in Fig. 4 (a). This is computed for different widths and radial TGs on the nanotrack. Here, $w_n$ is varied from 100nm to 300nm, and radial TG is varied from 5K/μm-degree to 20K/μm-degree for constant $\alpha_2 = 0.005$ and $\alpha_1 = 0.001$. The green circle represents that the skyrmion is able to perform uniform circular motion (UCM) on the nanotrack, meaning 'periodic' while the blue circle denotes that the skyrmion fails to complete the revolution, meaning 'non-periodic'. However, the red circle represents the 'annihilation' state where the skyrmion annihilates at the edges of the nanotrack.

The magnitude and direction of various forces that are acting on the skyrmion have different natures at various points including $J_1, J_2, H_1$ and $H_2$ on a nanotrack as shown in Fig. 4(b). For the force due to radial TG at different points, in order to provide the motion of skyrmion in CCW direction, $F_{TG\_net}$ should be greater than zero. However, in order to provide the CW motion, $F_{TG\_net}$ should be less than zero. As long as $F_{SkHE} = F_{sky-edge}$, the skyrmion will not be deformed, otherwise it would deform/annihilate. The magnitude and direction of forces acting on the skyrmion at different points ($J_1, J_2, H_1$ and $H_2$) on a nanotrack are responsible for three distinct behavior of skyrmion namely, periodic, non-periodic and annihilation that are listed in Table 2 and discussed in following sub-sections.

### A. Periodic

When $F_{sky-edge}$ and $F_{SkHE}$ are balanced in radial directions and $F_{TG\_net} > 0$, the skyrmion performs the UCM in CCW direction on a nanotrack with definite time period $T$. It is observed that when $150 \leq w_n \leq 300 nm$ and radial TG $\nabla T \leq 15$ K/μm-degree, the skyrmion has enough space on the nanotrack to stabilize itself at lower energy state (middle of nanotrack). The periodicity of the skyrmion can be seen from the instantaneous position of skyrmion in $xy$ plane that exhibits the definite frequency as shown in Fig. 4 (c).

### B. Non-periodic

When $F_{sky-edge}$ and $F_{SkHE}$ are balanced in radial directions but force due to radial TG acting on the skyrmion is not enough to make the skyrmion cross the energy step at the junction of two metal layers or intermediate position where $F_{TG\_net} > 0$ at the $J_1$ and $J_2$ and $F_{TG\_net} < 0$ at $H_1$ and $H_2$ when the width of nanotrack is not enough. It is observed that when the nanotrack width ranges from $100 \leq w_n < 150 nm$, the skyrmion is not able to perform the complete periodic motion as shown in Fig. 4(d).

Table 2. Illustration of forces acting on the different point on nanotrack during skyrmion motion

| Position on Nanotrack | Nanotrack width $w_n = R_2 - R_1$ for $\alpha_2 - \alpha_1 = 0.004$ and $\nabla T = 10$ K/μm-degree | | |
|---|---|---|---|
| | $100 \leq w_n < 150 nm$ | $150 \leq w_n < 250 nm$ | $250 \leq w_n < 300 nm$ |
| $H_2$ | $F_{net\_TG} < 0, F_{sky-edge} = F_{SkHE}$ | $F_{net\_TG} > 0, F_{sky-edge} = F_{SkHE}$ | $F_{net\_TG} < 0, F_{sky-edge} \neq F_{SkHE}$ |
| $J_2$ | $F_{net\_TG} > 0, F_{sky-edge} = F_{SkHE}$ | $F_{net\_TG} > 0, F_{sky-edge} = F_{SkHE}$ | $F_{net\_TG} > 0, F_{sky-edge} \neq F_{SkHE}$ |
| $H_1$ | $F_{net\_TG} > 0, F_{sky-edge} = F_{SkHE}$ | $F_{net\_TG} > 0, F_{sky-edge} = F_{SkHE}$ | $F_{net\_TG} > 0, F_{sky-edge} \neq F_{SkHE}$ |
| $J_1$ | $F_{net\_TG} < 0, F_{sky-edge} = F_{SkHE}$ | $F_{net\_TG} > 0, F_{sky-edge} = F_{SkHE}$ | $F_{net\_TG} < 0, F_{sky-edge} \neq F_{SkHE}$ |
| Skyrmion behaviour | 🔵 Non periodic | 🟢 Periodic | 🔴 Annihilation |

### C. Annihilation

Whenever $F_{sky-edge}$ and $F_{SkHE}$ are not balanced in radial directions and forces due to radial TG is high ($\nabla T \geq 15$ K/μm-degree), it tries to provide UCM of skyrmion on a nanotrack. However, due to higher temperature, the skyrmion doesn't has enough stability. The skyrmion gets accelerated and this leads to increment in gyromagnetic force that is not balanced by the force due to skyrmion-edge repulsion. Hence, the skyrmion is annihilated at the boundary as shown in Fig. 4(e).

## VII. WORKING WINDOW OF THE PROPOSED DEVICE FOR THE DAMPING CONSTANT DIFFERENCE

In this section, the working window for a circular motion of skyrmion on a nanotrack with respect to change in damping constants of two dissimilar nanotracks and radial TG is analysed and shown in fig. 5(a). However, the difference in the thermal field due to change in damping constant of nanotrack *A* and nanotrack *B* is shown in fig. 5(b). Here, $\alpha_2 - \alpha_1$ for the thermocouple nanotrack is varied from 0.002 to

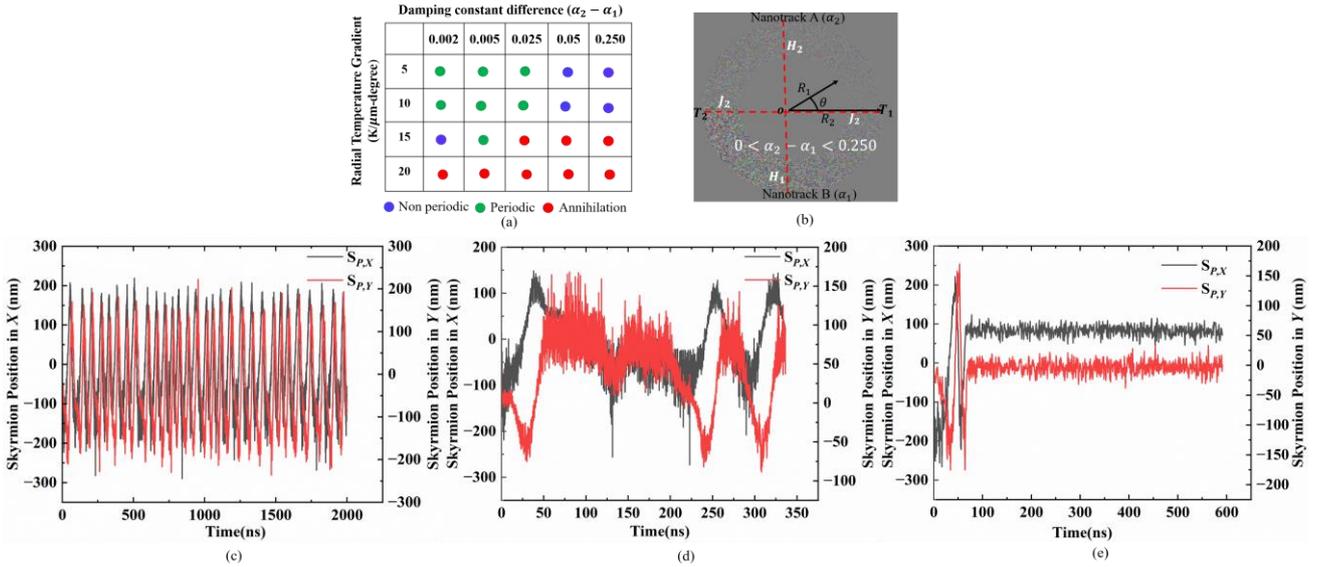

FIG. 5 (a) Working window for different skyrmion behaviour on a nanotrack width $w_n = 200 nm$ for $0.002 \leq \alpha_2 - \alpha_1 \leq 0.250$ and $5 \leq \nabla T \leq 20$ K/μm-degree (b) Thermal field in upper and lower part of nanotrack (c) Instantaneous skyrmion position when it exhibits periodic motion (d) Non-periodic motion and (e) Annihilation on a circular nanotrack.

0.250 and radial TG is varied from 5K/$\mu$m-degree to 20K/$\mu$m-degree. However, the width of the nanotrack is considered as $w_n = 200nm$. It is observed that variation in $\alpha_2 - \alpha_1$ play a vital role in the dynamics of skyrmion under radial TG.

### A. Periodic

For the case, when $0.002 < \alpha_2 - \alpha_1 \leq 0.025$, the skyrmion shows the periodic motion whenever radial TG ranges from 5K/$\mu$m-degree to 10K/$\mu$m-degree. This is due to the fact that when the difference between the damping constant is sufficient, the thermal spin current density ($T_2$ to $T_1$) is more in nanotrack $B$ as compared to the nanotrack $A$. Hence, the skyrmion move in CCW direction toward the $J_1$ through the point $H_1$ on nanotrack $B$ and it crosses the junction point $J_1$ due to the inertial effect and enters the nanotrack $A$. Here, the force caused by magnonic STT and induced dipolar field dominates over the force due to thermal STT, hence the skyrmion experiences $F_{TG\_net} > 0$ due to the fact that the magnonic wave propagate ($T_2$ to $T_1$) more effectively through nanotrack $A$ as compared to nanotrack $B$, hence the skyrmion move toward the $J_2$ through the point $H_2$ on the nanotrack $A$ and finally crosses the junction and reaches the nanotrack $B$ due to inertial effect. In this way, the complete periodic motion of the skyrmion is achieved with a definite time period as shown in Fig. 5 (c).

### B. Non-periodic

As the difference between the damping constant of both the nanotracks lie between $0.025 < \alpha_2 - \alpha_1 < 0.25$, the net effect of forces acting on the skyrmion is almost equal in both the nanotracks (very negligible change in forces due to magnonic STT and thermal STT acting on the skyrmion in the two nanotracks), meaning that $F_{TG\_net} > 0$ on nanotrack $B$ while $F_{TG\_net} < 0$ on nanotrack $A$. Hence, the force caused by magnonic STT and induced dipolar field is not able to overcome the force due to thermal STT that results in the pinning of the skyrmion on the nanotrack. Hence, it is observed that for $0.05 < \alpha_2 - \alpha_1 < 0.25$, the skyrmion is not able to perform any periodic motion on nanotrack for radial TG 5K/$\mu$m-degree to 10K/$\mu$m-degree as shown in Fig. 5(d).

### C. Annihilation

When the radial TG is high i.e. $10 \leq \nabla T \leq 15$ K/$\mu$m-degree on the nanotrack then the temperature of hotter end is reached upto 1020K and due to this, the skyrmion doesn't has enough stability. Therefore, during its motion, the skyrmion has large velocity on the nanotrack and this leads to the increment in gyromagnetic force that is not balanced by the force due to skyrmion-edge repulsion that results in instability of the skyrmion and finally it gets annihilated at the boundary as shown Fig. 5(e).

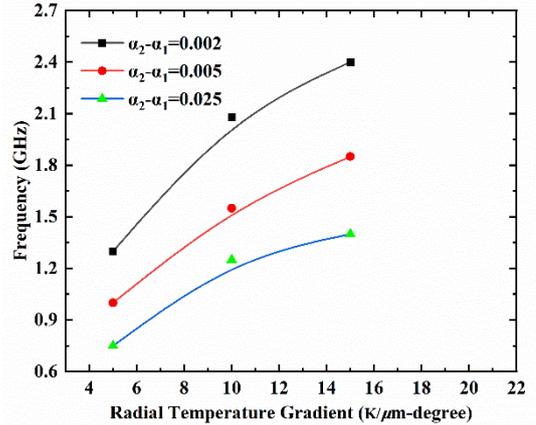

FIG. 6. The variation of frequency of skyrmion motion on the circular nanotrack for various values of difference of damping constants of upper and lower nanotracks. Here, radial TG is varying from 5K/$\mu$m-degree to 25K/$\mu$m-degree.

## VIII. FREQUENCY AND ENERGY ESTIMATION OF THE PROPOSED DEVICE

The variation of frequency of skyrmion motion at different $\alpha_2 - \alpha_1$ with radial TG varying from 5K/$\mu$m-degree to 25K/$\mu$m-degree is shown in Fig. 6. It is observed that the frequency of skyrmion motion increases with the radial TG however, the frequency is inversely proportional to $\alpha_2 - \alpha_1$. This is due to the fact that the larger value of radial TG and lower value of damping constant provide high speed revolution of skyrmion that further results in increase in the frequency of skyrmion motion. In this analysis, at $\alpha_2 - \alpha_1 \sim 0.002$ and radial TG=15K/$\mu$m-degree, the maximum frequency is observed that is estimated as 2.5 GHz.

The thermal energy required to maintain the temperature difference at two ends can be expressed as: $E_{th} = k_2 A_0 (\nabla T) t_2 + k_1 A_0 (\nabla T) t_1 = 0.84 fJ$/oscillation, where $k_2 = 2.23 W m^{-1} K^{-1}$ and $k_1 = 2 W m^{-1} K^{-1}$ are the thermal conductivities corresponding to $\alpha_2$ and $\alpha_1$ [40]. Here, $t_2$ and $t_1$ are the time period of skyrmion motion on nanotrack $A$ and $B$, respectively. The value of cross-sectional area ($A_0$), $\partial T$, $w_n$ and $\partial \theta$ is considered as $200 \times 2 nm^2$, 256K, $200nm$ and 180°, respectively. However, the thermal energy can be avoided if the waste heat is used to create the gradient on the nanotrack. These advantages of the proposed device could further minimize the energy consumption of emerging further spintronic devices.

## IX. CONCLUSION

TG is one of the most promising alternative for generating driving power for spintronic-based devices in an energy-efficient manner. The skyrmion motion on a thermocoupled nanotrack is investigated for the first time. The thermocoupled nanotrack is created by connecting two nanotracks A and B, end to end that have different damping constants $\alpha_2$ and $\alpha_1$, respectively. This has been done in order to set up a spontaneous pure spin current on a thermocoupled nanotrack, where a difference in temperature is maintained between the ends. Later, an oscillator that generates thermal driving power for a skyrmion on a thermocoupled nanotrack is proposed. It is investigated that the width of the nanotrack and the difference in the damping constant between the both nanotracks play a vital role in achieving the proper periodic motion of the skyrmion on the circular nanotrack. If the width of nanotrack will be too narrow, the energy well will not be created and the skyrmion will not perform periodic motion. However, if the width is too high, the skyrmion is annihilated at the boundary due to unbalancing of forces from SkHE and edge repulsion. If the radial TG is too large $\nabla T \geq 15$ K/μm-degree then the skyrmion gets annihilated almost in all cases. However, It is observed that for the skyrmion to attain the sustainable periodic motion, the width of the nanotrack and difference in the damping constants should be considered in the range $150 \leq w_n \leq 250 nm$ and $0.002 < \alpha_2 - \alpha_1 < 0.025$, respectively. The maximum frequency observed is 2.5 GHz and it is achievable at $\alpha_2 - \alpha_1 \sim 0.002$. Moreover, the energy consumption for driving the skyrmion on nanotrack is 0.84fJ/oscillation hence, this opens up the path for the development of energy-efficient skyrmion based devices.

## ACKNOWLEDGEMENTS

R. K. Raj, N. Bindal, and B. K. Kaushik would like to acknowledge Council of Scientific and Industrial Research (CSIR) Grant No. 09/0143(11108)/2021-EMR-I, Science and Engineering Research Board (SERB), Department of Science and Technology, Government of India (GOI) under Grant CRG/2019/004551 for providing the funding to carry out the research work.

## REFERENCES

[1] N. Nagaosa and Y. Tokura, Topological properties and dynamics of magnetic skyrmions, Nat. Nano. **8,** 899 (2013).
[2] Q. Zhang, J. Liang, K. Bi, L. Zhao, H. Bai, Q. Cui, H.A. Zhou, H. Bai, H. Feng, W. Song, and G. Chai, Quantifying the Dzyaloshinskii-Moriya interaction induced by the bulk magnetic asymmetry, Phys. Rev. Lett. **128,** 167202 (2022).
[3] A. Soumyanarayanan, M. Raju, A.L. Gonzalez Oyarce, M.Y. Petrović, A.P. Ho, P. Khoo, K.H. Tran, M. Gan, and F. Ernult, Tunable room-temperature magnetic skyrmions in Ir/Fe/Co/Pt multilayers, Nat. Mat. **16**, 898 (2017).
[4] M. Xin, Y. Guoqiang, L. Xiang, W. Tao, W. Di, K. S. Olsson, C. Zhaodong, A. Kyongmo, Q. X. John, L. W. Kang, and L. Xiaoqin, Interfacial control of Dzyaloshinskii-Moriya interaction in heavy metal/ferromagnetic metal thin film heterostructures, Phys. Rev. B **94**, 180408 (2016).
[5] G. Yu, A. Jenkins, X. Ma, S. A. Razavi, C. He, G. Yin, Q. Shao, Q. l. He, H. Wu, W. Li, W. Jiang, X. Han, X. Li, A. B. Jayich, P. K. Amiri, and K. L. Wang, Room-temperature skyrmions in an antiferromagnet-based heterostructures, Nano Lett. **18**, 980 (2018).
[6] A. Saini, A. Dwivedi, A. Lodhi, Khandelwal, and S. P. Tiwari, Resistive switching behaviour of TIO2/ (PVP: MoS2) nanocomposite hybrid bilayer in rigid and flexible RRAM devices, Memories - Materials, Devices, Circuits and Systems, **4**, 100029 (2023).
[7] S. Luo, and L. You, Skyrmion devices for memory and logic applications, APL Mat. **9,** 050901 (2021).
[8] X. Zhao, R. Ren, G. Xie, and Y. Liu, Single Antiferromagnetic skyrmion transistor based on strain manipulation, Appl. Phys. Lett. **112**, 252402 (2018).
[9] N. Bindal, R. K. Raj, and B. K. Kaushik, Antiferromagnetic skyrmion-based high speed diode, Nano. Adv. **5,** 450 (2023).
[10] G. Yu, X. Xu, Y. Qiu, H. Yang, M. Zhu, and H. Zhou, Strain-modulated magnetization precession in skyrmion-based spin transfer nano-oscillator, Appl. Phys. Lett. **118**, 142403 (2021).
[11] N. Bindal, R. K. Raj, and B. K. Kaushik, Antiferromagnetic skyrmion based shape-configured leaky-integrate-fire neuron device, J. Phys. D: Appl. Phys. **55,** 345007 (2022).
[12] Z. Zhongming, G. Finocchio, and H. Jiang, Spin transfer nano-oscillators, Nano. **5,** 2219 (2013).
[13] R. Cheng, D. Xiao, and B. Arne, Terahertz antiferromagnetic spin Hall nano-oscillator, Phys. Rev. Lett. **116**, 207603 (2016).
[14] D. Das, B. Muralidharan, and A. Tulapurkar, Skyrmion based spin-torque nano-oscillator, J. Magn. Magn. **491,** 165608 (2019).
[15] J. H. Guo, J. Xia, X. C. Zhang, P. W. T. Pong, Y. M. Wu, H. Chen, W. S. Zhao, and Y. Zhou, A ferromagnetic skyrmion-based nano-oscillator with modified profile of Dzyaloshinskii-Moriya interaction, J. Magn. Magn. **496,** 165912 (2020).
[16] D. Bhattacharya and J. Atulasimha, Skyrmion-mediated voltage-controlled switching of ferro-magnets for reliable and energy-efficient two-terminal memory, ACS Appl. Mater. Interfaces **10**, 17455 (2018).
[17] Z. Wang, M. Guo, H. Zhou, L. Zhao, T. Xu, R. Tomasello, H. Bai, Y. Dong, S. Je, W. Chao, H. Han, S.Lee, K. Lee, Y. Yao, W.Han, C. Song, H. Wu, M. Carpentieri, G. Finocchio, M. Im, S. Lin, and W. Jiang, Thermal generation, manipulation and thermoelectric detection of skyrmions, Nat. Electron **3**, 672 (2020).
[18] G. Qin, X. Zhang, R. Zhang, K. Pei, C. Yang, C. Xu, Y. Zhou, Y. Wu, H. Du, and R. Che, Dynamics of magnetic skyrmions sriven by a temperature gradient in a chiral magnet FeGe, Phys. Rev. B **106** 024415 (2022).
[19] C. Gong, Y. Zhou, and G. Zhao, Dynamics of magnetic skyrmions under temperature gradients, Phys. Rev. Lett. **120** (2022)**.**
[20] E. Raimondo, E. Saugar, J. Barker, D. Rodrigues, A. Giordano, M. Carpentieri, W. Jiang, O. Chubykalo-Fesenko, R. Tomasello, and G. Finocchio, Temperature gradient-driven magnetic skyrmion motion, Phys. Rev. Appl. **18,** 024062 (2022).
[21] L. Kong and J. Zang, Dynamics of an insulating skyrmion under a temperature gradient, Phys. Rev. Lett. **111,** 067203 (2013).
[22] X. Yu, F. Kagawa, S. Seki, M. Kubota, J. Masell, F. S. Yasin, K. Nakajima, M. Nakamura, M. Kawasaki, N.Nagaosa and Y. Tokura Real-space observations of 60-nm skyrmion dynamics in an insulating magnet under low heat flow, Nat. Commun. **12**, 5079 (2021).
[23] F. Schlickeiser, U. Ritzmann, D. Hinzke, and U. Nowak, Role of entropy in domain wall motion in thermal gradients, Phys. Rev. Lett. **113**, 097201 (2014).
[24] A. Vansteenkiste, J. Leliaert, M. Dvornik, M. Helsen, F. G. Sanchez, and B. V. Waeyenberge, The design and verification of Mumax3, AIP Adv. **4**, 10713 (2014).

[25] J. Leliaert, M. Dvornik, J. Mulkers, J. D. Clercq, M. V. Milošević, and B. V. Waeyenberge, Fast micromagnetic simulations on GPU-recent advances made with mumax3, J. Phys. D: Appl. Phys. **51**, 123002 (2018).

[26] J. Q. Lin, J. P. Chen, Z. Y. Tan, Y. Chen, Z. F. Chen, W. A. Li, X. S. Gao, and J. M. Liu, Manipulation of skyrmion motion dynamics for logical device application mediated by inhomogeneous magnetic anisotropy, Nano. Mat. **12**, 278 (2022).

[27] C. C. I. Ang, W. Gan, and W. S. Lew, Bilayer skyrmion dynamics on a magnetic anisotropy gradient, New J. Phys. **21,** 043006 (2019).

[28] J. sampaio, V. Cros, S. Rohart, A. Thiaville, and A. Fert, Nucleation, stability and current-induced motion of isolated magnetic skyrmions in nanostructures, Nat. Nano. **8**, 893 (2013).

[29] M. T. Islam, X. S. Wang, and X. R. Wang, Thermal gradient driven domain wall dynamics, J. Condens. Mat. Phys. **31**, 455701 (2019).

[30] V. Puliafito, R. Khymyn, M. Carpentieri, B. Azzerboni, V. Tiberkevich, A. Slavin, and G. Finocchio, Micromagnetic modeling of terahertz oscillations in an antiferromagnetic material driven by the spin Hall effect, Phys. Rev. B **99**, 024405 (2019).

[31] J. Flipse, F. L. Bakker, A. Slachter, F. K. Dejene, and B. J. Wees, Direct observation of the spin-dependent Peltier effect, Nat. Nano. **7,** 166 (2012).

[32] S. Meyer, Y. T. Chen, S. Wimmer, M. Althammer, T. Wimmer, R. Schlitz, S. Geprägs, H. Huebl, D. Ködderitzsch, H. Ebert, G. E. W. Bauer, R. Gross, and S. T. B. Goennenwein, Observation of spin Nerst effect, Nat. Mater. **16**, 977 (2017).

[33] S. Park, N. Nagaosa, and B. J. Yang, Thermal Hall effect, spin Nerst effect, and spin density induced by a thermal gradient in collinear ferrimagnets from magnon-phonon interaction, Nano Lett. **20**, 2741 (2020).

[34] K. Uchida, S. Takahashi, K. Harii, J. Ieda, W. Koshibae, K. Ando, S. Maekawa, and E. Saitoh, Observation of the spin Seebeck effect, Nat. Phys. **455**, 778 (2008).

[35] X. Zhang, J. Müller, J. Xia, M. Garst, X. Liu, and Y. Zhou, Motion of Skyrmions in nanowires driven by magnonic momentum-transfer forces, New J. Phys. **19**, 065001 (2017).

[36] H. Chang, P. A. Praveen Janantha, J. Ding, T. Liu, K. Cline, J. N. Gelfand, W. Li, M. C. Marconi, and M. Wu, Role of damping in spin Seebeck effect in yttrium iron garnet thin films, Sci. Adv. **3**, 1601614 (2017).

[37] J. Cramer, U. Ritzmann, B.-W. Dong, S. Jaiswal, Z. Qiu, E. Saitoh, U. Nowak, and M. Kläui, Spin transport across antiferromagnets induced by the spin Seebeck effect, J. Phys. D: Appl Phys. **51**, 14 (2018).

[38] M. Weißenhofer, L. Rózsa, and U. Nowak, Skyrmion dynamics at finite temperatures: Beyond Thiele's equation, Phys. Rev. Lett. **127**, 047203 (2021).

[39] X. S. Wang, H. Y. Yuan and X. R. Wang, A theory on skyrmion size, Commun. Phys. **1,** 31 (2018).

[40] R. Das, A. Chanda, and R. Mahendiran, Influence of magnetic field on electrical and thermal transport in the hole doped ferromagnetic manganite: $La_{0.9}Na_{0.1}MnO_3$, RSC Adv. **9**, 1726 (2019).